\documentclass[sigconf]{acmart}
\usepackage{savesym}

\savesymbol{pdfbookmark}
\usepackage{etoolbox}

\savesymbol{pdfbookmark}
\usepackage{latexsym}

\savesymbol{pdfbookmark}
\usepackage{tcolorbox}

\savesymbol{pdfbookmark}
\usepackage{mdframed}

\savesymbol{pdfbookmark}
\usepackage{url}

\savesymbol{pdfbookmark}
\usepackage{xcolor}
\def\shortname{\textsc{SEAL}}
\newcommand\BibTeX{B\textsc{ib}\TeX}

\AtBeginDocument{%
  \providecommand\BibTeX{{%
    \normalfont B\kern-0.5em{\scshape i\kern-0.25em b}\kern-0.8em\TeX}}}

\setcopyright{rightsretained}

\begin{document}

\copyrightyear{2020}
\acmYear{2020}

\acmConference[JCDL '20]{Proceedings of the ACM/IEEE Joint Conference on Digital Libraries in 2020}{August 1--5, 2020}{Virtual Event, China}
\acmBooktitle{Proceedings of the ACM/IEEE Joint Conference on Digital Libraries in 2020 (JCDL '20), August 1--5, 2020, Virtual Event, China}
\acmDOI{10.1145/3383583.3398625}
\acmISBN{978-1-4503-7585-6/20/06}

\fancyhead{}
\title{\shortname{}: Scientific Keyphrase Extraction and Classification}

\author{Ayush Garg, Sammed Shantinath Kagi, Mayank Singh}
\email{{ayush.g,sammed.shantinath,singh.mayank}@iitgn.ac.in}
\affiliation{%
 \institution{Indian Institute of Technology, Gandhinagar}
 }

\begin{abstract}
Automatic scientific keyphrase extraction is a challenging problem facilitating several downstream scholarly tasks like search, recommendation, and ranking. In this paper, we introduce \shortname{}, a scholarly tool for automatic keyphrase extraction and classification. The keyphrase extraction module comprises two-stage neural architecture composed of Bidirectional Long Short-Term Memory cells augmented with Conditional Random Fields. The classification module comprises of a Random Forest classifier. We extensively experiment to showcase the robustness of the system. We evaluate multiple state-of-the-art baselines and show a significant improvement. The current system is hosted at \url{http://lingo.iitgn.ac.in:5000/}.
\end{abstract}


\maketitle

\section{Introduction}
With the ever-growing scientific volume, scholarly search and recommendation engines are gradually adopting artificial intelligence frameworks for better document retrieval. A well-known annotation task is to identify topical keyphrases for facilitating topical search. Further,  the identified keyphrases can be classified into several semantic categories for facilitating knowledge graph construction.  However, due to the large volume of research, current manual annotation schemes are financially infeasible owing to the requirement of continuous human resources and domain expertise. 

In this paper, we introduce \shortname{} that aims to automate keyphrase extraction and further classification into three semantic categories: (i) tasks, (ii) processes, or (iii) materials. Tasks represent research problems like extraction, processing, parsing, etc. Processes represent solutions to problems, including physical equipment, algorithms, methods/techniques, and tools. Materials include physical material such as chemical compounds and datasets. We showcase that \shortname{} outperforms several state-of-the-art tools on a recently published dataset of 500 scientific publications in the field of Computer Science, Material Sciences, and Physics~\cite{AugensteinDRVM17}.

\section{\shortname{} Architecture}
\begin{figure*}
\vspace{-1cm}
    \centering
    \begin{tabular}{cc}
      \includegraphics[width=0.5\hsize]{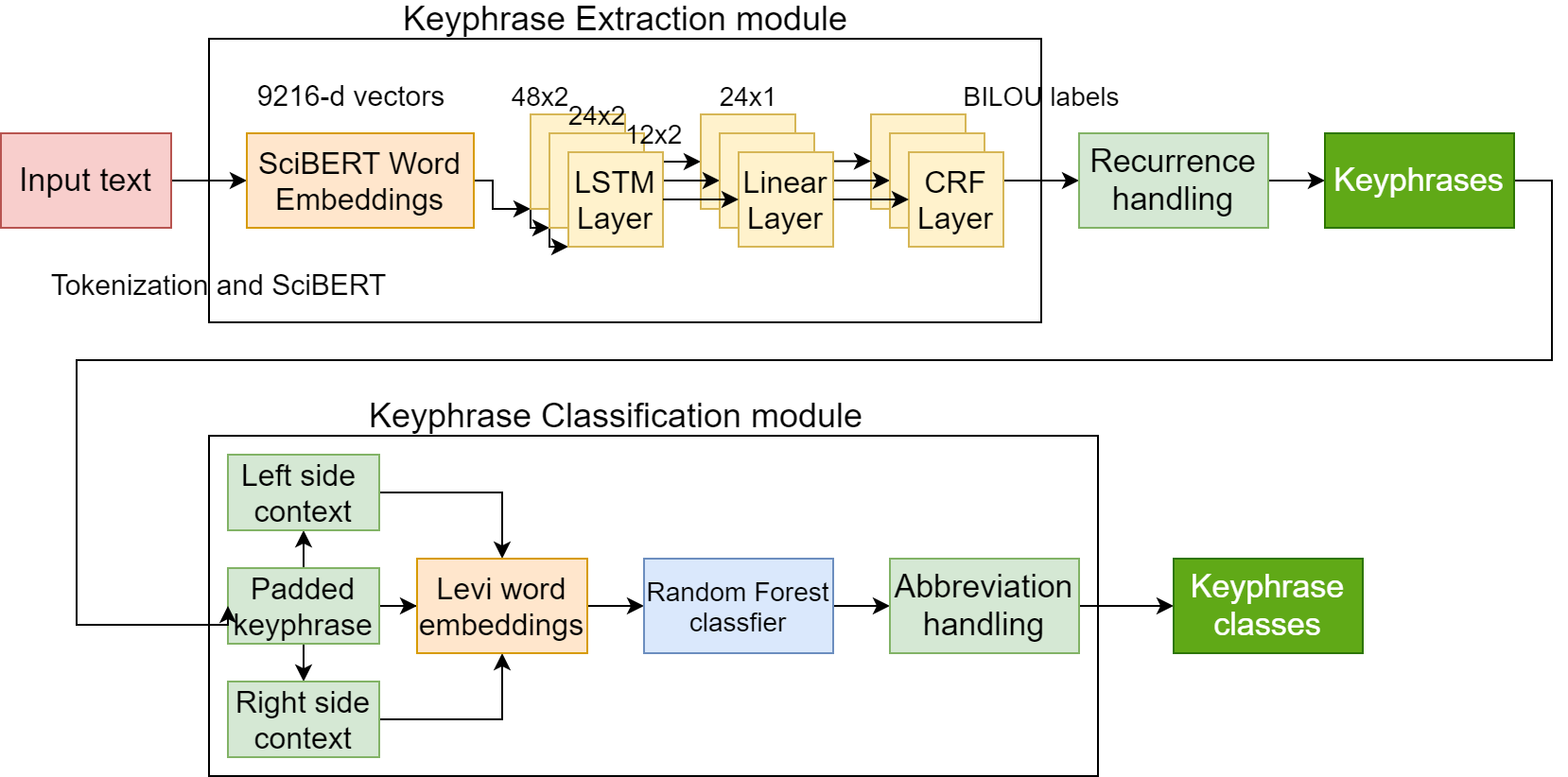} & \includegraphics[width=0.5\hsize]{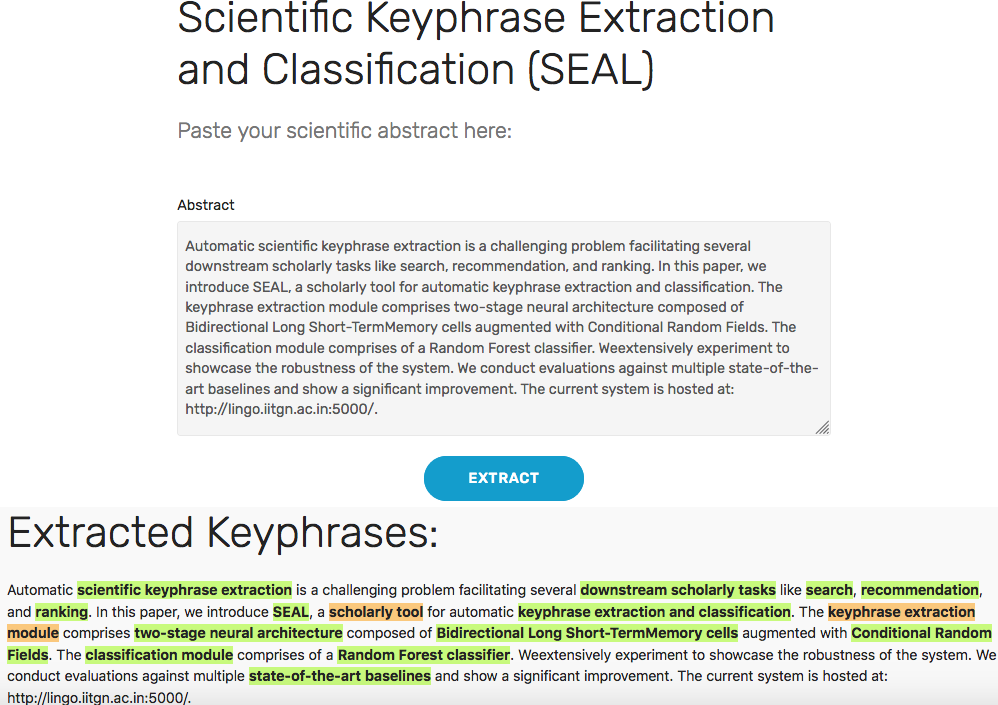} \\
      (a) & (b)
    \end{tabular}
    \vspace{-0.3cm}
    \caption{ (a) Flow chart of the \shortname{} architecture. (b) Snapshot of SEAL demo webpage with the output of abstract of this paper. Processes, materials and tasks are marked in green, orange and blue respectively.}\label{fig:architecture}
\vspace{-0.5cm}
\end{figure*}

\shortname{} comprises two distinct neural modules, one for keyphrase extraction and other for keyphrase classification. We use standard \textit{`Beginning, Inside and Last tokens of multi-token chunks, Unit-length chunks and Outside'} (BILOU) labeling scheme~\cite{ammar-etal-2017-ai2} in both of the modules. Figure~\ref{fig:architecture}(a) presents the \shortname{} architecture.
\subsection{Keyphrase Extraction Module}
This module leverages pre-trained token-level 9216-dimensional SciBERT embeddings \cite{abs-1903-10676}\footnote{SciBERT results in significantly better scores than several other embeddings such as Levi and Goldberg dependency-based embeddings~\cite{levy-goldberg-2014-dependency} and GLOVE~\cite{pennington-etal-2014-glove}.} to train three layers (96,48, and 24 hidden units in respective layers) of Bidirectional Long Short-Term Memory (BiLSTM) cells stacked on top of each other.  The output, a 24-dimensional vector, is then downsized to a five-dimensional vector using a linear layer and then fed to a Conditional Random Field (CRF) layer, which then predicts the label of the token. The results are, further, refined through a post-processing step to handle single-token keyphrases.

\subsection{Keyphrase Classification Module}
This module uses pre-trained token-level Levy Embeddings~\cite{levy-goldberg-2014-dependency}\footnote{Levy embeddings results in significantly better scores than several other embeddings.}. For each token, we also consider the immediate neighboring tokens as context. We pass the concatenated vector of the embedding of the current token, the previous token, and the next token to the standard Random Forest (RF) classifier. Candidate tokens without next or previous tokens are appropriately padded with the embedding corresponding to the \textit{<UNKNOWN>} tag. 

We post-process the abbreviations and chemical formulae separately due to inefficiencies associated with their classification. In the case of abbreviations, we match abbreviations with the full-form using their first occurrence and assigned them the respective class of the full-form. In the case of chemical formulae (such as NaCl, Mg), we match the corresponding formulae name token using a regular expression.

\section{Experimental Results}
We experiment on ScienceIE dataset\footnote{https://scienceie.github.io/}~\cite{AugensteinDRVM17} containing 500 scientific abstracts curated from Science Direct open access publications. Each abstract is manually labeled with keyphrase boundaries and their respective classes. For experimentation, we verbatim follow the guidelines specified in the ScienceIE competition. The dataset is partitioned into train, development, and test sets containing 350, 50, and 100 abstracts, respectively. We next, showcase that \shortname{} outperformed the top-ranked implementations on the ScienceIE leaderboard against the standard F1-score metric.

\noindent \textbf{Keyphrase Extraction:} 
As described in previous section, we experiment with several embedding schemes. Table~\ref{tab:embeddings} shows extraction accuracy of SciBERT outperformed other embedding schemes. Table~\ref{tab:embeddings} also compares F1-scores of \shortname{} against the ScienceIE official leaderboard top-rankers~\cite{AugensteinDRVM17}. Furthermore, unlike TIAL\_UW (rank 1 in the leaderboard ) and s2\_end2end~\cite{ammar-etal-2017-ai2} (rank 2 in the leaderboard), \shortname{} does not use external data sources. 

\noindent \textbf{Keyphrase Classification}
Table~\ref{tab:classfication} compares F1-scores of \shortname{} against the ScienceIE official leaderboard~\cite{AugensteinDRVM17}. Note that, classification module leverages Levy embeddings.

\begin{table}[!tbh]
\begin{tabular}{p{5.5cm} p{2cm}}
\toprule
Model & F1 score \\
\midrule
Glove word embeddings \cite{pennington-etal-2014-glove} & 0.440 \\
Levi and Goldberg embeddings \cite{levy-goldberg-2014-dependency}   & 0.470 \\
SciBERT embeddings \cite{abs-1903-10676}   & 0.564\\
\midrule
\shortname{} & \textbf{0.564} \\
TIAL\_UW & 0.560 \\
s2\_end2end \cite{ammar-etal-2017-ai2}   & 0.550 \\
\bottomrule
\end{tabular}
\caption{Performance of \shortname{} at different input embeddings and against state-of-the-art extraction systems described in ScienceIE leaderboard~\cite{AugensteinDRVM17}.}\label{tab:embeddings}
\vspace{-0.6cm}
\end{table}

\begin{table}[!tbh]
\begin{tabular}{p{5.5cm} p{2cm}}
\toprule
Model & F1-score \\
\midrule
\shortname{} & \textbf{0.74} \\
MayoNLP \cite{liu-etal-2017-mayonlp}& 0.67 \\
UKP/EELECTION  \cite{eger-etal-2017-eelection} & 0.66 \\
\bottomrule
\end{tabular}\caption{Performance of \shortname{} against  top-ranked classification systems described in ScienceIE leaderboard~\cite{AugensteinDRVM17}.}\label{tab:classfication}
\vspace{-0.5cm}
\end{table}

\section{System Description}

The \shortname{} web-application is developed using the Flask framework. The current implementation uses Pytorch Framework for the extraction and classification modules. The trained models and the demo are hosted at our research group server\footnote{\url{http://lingo.iitgn.ac.in:5000/}}. Figure~\ref{fig:architecture}(b) presents snapshot of the demo. On encountering a POST request, the framework first executes the extraction module, followed by the classification module, and displays the result. The code, processed dataset, and the system implementation details are available at \url{https://github.com/Sammed98/Keyphrase-Extraction-Demo}.

\section{Conclusion and future proposals}
In this paper, we propose a toolkit \shortname{} for scientific keyphrase extraction as well as classification. We showcase that \shortname{} performed similar to state-of-the-art extraction systems that leverage the large volume of external knowledge. In the future, we plan to experiment with domain-specific embeddings and semi-supervised bootstrapping techniques.

\bibliographystyle{ACM-Reference-Format}

\end{document}